\input  phyzzx
\input epsf
\overfullrule=0pt
\hsize=6.5truein
\vsize=9.0truein
\voffset=-0.1truein
\hoffset=-0.1truein

%
%
\def\half{{1\over 2}}
\def\IC{{\ \hbox{{\rm I}\kern-.6em\hbox{\bf C}}}}
\def\IR{{\hbox{{\rm I}\kern-.2em\hbox{\rm R}}}}
\def\IZ{{\hbox{{\rm Z}\kern-.4em\hbox{\rm Z}}}}
\def\pa{\partial}
\def\sIR{{\hbox{{\sevenrm I}\kern-.2em\hbox{\sevenrm R}}}}

\def\sym{super Yang Mills theory}

%
%
\hyphenation{Min-kow-ski}

\rightline{SU-ITP-97-11}
\rightline{April  1997}
\rightline{hep-th/9704080}

\vfill

%
%
\title{Another Conjecture about M(atrix) Theory}

\vfill

%
%

\author{Leonard Susskind\foot{susskind@dormouse.stanford.edu}}

\vfill

\address{Department of Physics \break Stanford University,
 Stanford, CA 94305-4060}

\vfill

%
%

The current understanding of M(atrix) theory is that in the
large N limit certain supersymmetric Yang Mills theories become  
equivalent to
M-theory in the infinite momentum frame.  In this paper  the  
conjecture is put
forward that  the equivalence between M and M(atrix) theory is not  
limited to
the large N limit but  is valid for finite N.  It is argued that a  
light cone
description of M-theory exists in which one of the light like  
coordinates is
periodically identified. In the light cone literature this is  
called Discrete
Light Cone Quantization (DLCQ). In this framework an exact light cone
description exists  for each quantized  value N of longitudinal  
momentum.  The
new conjecture states that the sector of  the DLCQ   of M-theory is  
exactly
described by a U(N) matrix theory. Evidence is presented for the
conjecture.
\vfill\endpage

%
%
\REF\bfss{T. Banks, W. Fischler, S. Shenker and L.  
Susskind,hep-th/9610043.}
\REF\imf {   L.Susskind, Boulder Summer School Lectures (1968)}
\REF\DLCQ{H.C. Pauli and S.J. Brodsky,Phys.Rev.{\bf D32} (1985) 1993.}

\REF\m{L. Motl, hep-th/9701025.}
\REF\bs{T. Banks and N. Seiberg, hep-th/9702187.}
\REF\dvv{R. Dijkgraaf, E. Verlinde and H. Verlinde, hep-th/9703030.}
\REF\gs{M. Green and  N. Seiberg Nucl. Phys. B299 :559, 1988}
\REF\kp{D. Kabat and P. Pouliot,``A
Comment on Zero-Brane Quantum Mechanics", hep-th/9603127}
\REF\dfs{U.H. Danielsson, G. Ferretti and B. Sundborg,``D-particle
Dynamics and Bound States", hep-th/9603081}
\REF\wati{W.Taylor, hep-th/9611042.}
\REF\s{L. Susskind, hep-th/9611164.}
\REF\grt{O.J. Ganor, S. Ramgoolam and W. Taylor,hep-th/9611202.}
\REF\fhrs{W. Fishler, E. Halyo, R. Rajaraman and L. Susskind,  
hep-th/9703102 }
\REF\ms{J. Maldacena and L. Susskind, Nucl. Phys. {\bf B475} (1996) 670,
hep-th/9604042.}
\REF\ss{S. Sethi and L. Susskind, hep-th/9702101.}
\REF\motl{Similar arguments have also been made by L. Motl,
private communication.}
\Ref\p{Polchinski and Pouliot have given evidence that full Lorentz  
invariance
is restored in the large N limit. Joseph Polchinski and Philippe Pouliot,
 Membrane Scattering with M-Momentum Transfer  hep-th/9704029.}



%
%

%
%
\chapter{Introduction}

According to our present understanding of M(atrix) theory,
the connection between supersymmetric matrix field theories   and  
M-theory
is only  precise in the large N limit where it becomes the infinite
momentum limit  of M-theory.  No exact significance is given to the  
matrix
model for finite N.  In this paper arguments will be given for  an exact
connection  even at finite N.

In [\bfss] the tool that was used  to relate M and M(atrix) theory  
was the
Infinite Momentum Limit [\imf]. The theory is first compactified in  
a spacelike
direction $X^{11}$ with compactification radius $R$. The momentum  
$p_{11}$ is
quantized in units of $1/R$. Thus an integer $N=p_{11} R$ is  
defined. It is
then argued that In the $N \to \infty$ objects with vanishing and
negative $p_{11}$ decouple.  Since the only objects in IIA string  
theory which
carry  $p_{11}$ are the D0-branes, M-theory in the Infinite  
Momentum Limit
must be the theory of N D0-branes in the limit of large N.

The Infinite momentum method and the method of light cone  
quantization are
usually considered to be the same thing. However there is a subtle  
difference
especially when the longitudinal direction is compactified.   In  
the method
of Discrete
Light Cone Quantization (DLCQ) [\DLCQ] the coordinate which is  
compactified
is not the
space-like
$X^{11}$ but rather the lightlike coordinate $X^{-}$. In this case the
discrete momentum
is $p_{-}$ which is quantized as $p_{-} = {N \over R}$. In the  
limit $N \to
\infty$ the  the IML and DLCQ are expected to become identical.   
For finite N
the two are different. For example when N is finite, negative and  
vanishing
$p_{11}$ does not decouple, nor does the system have Galilean  
invariance. By
contrast in DLCQ the Galilean invariance, decoupling of negative  
$p_{-}$ and
the simplicity of the vacuum is exact for all N. In this paper the  
proposal is
put forward that the DLCQ of M-theory is exactly described by  
$U(N)$ super
Yang Mills theory.

The evidence for the conjecture is of two kinds.  Both involve the
transversely compactified versions of the theory. The first piece  
of evidence
involves the compactification on a circle in the limit of vanishing  
radius. In
this limit M-theory becomes free type IIA
 string theory. As we shall see this
theory admits DLCQ and its spectrum is easily worked out.
 Furthermore,  following the work of Motl [\m], Banks and Seiberg  
[\bs] and
Dijkgraaf, Verlinde and
Verlinde [\dvv] we find that the spectrum agrees with that of super Yang
Mills theory in the
appropriate limit. Furthermore there is strong evidence from [\dvv]  
that the
agreement  is not limited to the free string limit.

The second kind of evidence has to do with the constraints imposed   
on the
super Yang Mills theory by the  U-duality of M-theory.  In the original
proposal of [\bfss] there was no particular reason why these  
conditions should
be satisfied other than at infinite N. In the present interpretation the
dualities  should be correct for all N.  That meshes very well with  
the fact
that  \sym \ has the required dualities for finite N.

The plan of the paper is as follows. In section 2 a brief  
explanation of the
difference between the infinite momentum frame  and the light cone frame
 is given.

In section 3 free string theory is formulated in the light cone  
frame with the
lightlike coordinate $X^{-}$ compactified on a circle of radius  
$R$. As usual,
periodic identification of a coordinate entails two steps. The  
first step is to
quantize the conjugate momentum $p_{-}$ in units of $1/R$. The  
second step is
to introduce twisted sectors describing strings which are wound  
around the
compact direction.  The theory consists of sectors in which the  total
momentum $p_{-}$  is set equal to $N/R$. The spectrum is easily computed.

M(atrix) theory, in the limit in which a transverse coordinate is  
compactified
to zero size is expected to yield type IIA string theory. The  
transversely
compactified theory is described by strongly coupled 1+1  
dimensional \sym \
with gauge group $U(N)$. Using the method of  [\m] [\bs] [\dvv] we
find that
the spectrum of this field theory exactly matches the spectrum of  
the string
theory at momentum $p_{-} =N/R$. It is also possible to see that  
perturbations
away from the free string theory are in agreement with the strong  
coupling
expansion of the \sym.

In section 4 we review the constraints on \sym \ implied by  the  
dualities of
M-theory. From the point of view of [\bfss] these constraints are   
required to
be true only in the limit $N \to \infty$. What we find is that they  
are true
for
all N.  This fact now finds a natural interpretation in the context  
of DLCQ.

%
%
\chapter{Light  Cone Versus Infinite Momentum }

The  infinite momentum  description   given in [\bfss] begins by  
considering a
system of particles in a conventional reference frame.  The energy of a
collection of free particles is
$$
E=\sum_{a} \sqrt{p_a^2 + m_a^2}
\eqn\twoone
$$
where $p_a$ is the 3-momentum of the $a$th particle and $m_a$ is  
its mass.
The system is now boosted along the spatial ``longitudinal" direction
$X^{11}$ until the longitudinal momentum of every particle is  
positive and
much larger than any
other mass scale in the problem. We will call such a state a
``proper" state. In this limit  the energy takes the form
$$
E=p_{11}(total) + \sum_{a} {P^2 + m^2 \over 2 p_{11}}
\eqn\twotwo
$$
where $p_{11}(total)$ is the total longitudinal momentum and $P$
 stands for the transverse spatial momentum of a particle.
Note that the typical energy difference between two states of the  
same total
longitudinal momentum goes to zero like $1/p_{11}$. The leading  
term in $E$ can
be dropped for most purposes since it does not influence energy  
differences.

Let us consider a state with total longitudinal momentum  
$p_{11}(total)$ but
which happens to contain a particle with negative $p_{11}$ of the  
same order
as  $p_{11}(total)$. Such an ``improper" state   differs in energy  
from proper
states by a large amount of order  $p_{11}(total)$. For this reason as
$p_{11}(total) \to \infty$, improper states with backward moving quanta
decouple  in the IMF.

Some features of quantization in the IMF make it especially  
attractive [\imf].
\item{1.} Physics in the IMF has a galilean invariance which makes its
structure similar to that of nonrelativistic quantum mechanics. Thus the
concepts of  wave function, bound state and mass conservation have  
their naive
nonrelativistic meaning. Eq \twotwo \ is an example of this.

\item{2.} All objects carrying negative $p_{11}$ decouple for the  
reason just
stated.

\item{3.} In the limit of infinite momentum the complex structure of the
vacuum decouples from all proper systems. This means that for all  
practical
purposes, the vacuum is trivial as it is  in nonrelativistic quantum
mechanics. None of the these simplifications take place  as long as  
$p_{11}$ is
not essentially infinite.

In [\bfss] the longitudinal spacelike coordinate $X^{11}$ was  
assumed to be
compactified on a large circle of radius $R$ which is allowed to become
infinite eventually. However for finite $R$ the effect of the
compactification is to quantize the momentum $p_{11}$ in integer  
multiples of
$1/R$. Thus a quantum number $N$ is defined
$$
N=p_{11}(total) R
\eqn\twothree
$$
Similarly the momentum of any constituent subsystem $a$ satisfies
$$
N(a)=p_{11}(a) R
\eqn\twofour
$$
All of the simplifications described above still apply but only in  
the limit
$N\to \infty$.

Now let us consider a different approach which is usually called  
light cone
quantization. To make the discussion concrete, we will work out the light
cone quantization of $\Phi^3$ theory. The action is given by
$$
I=\int d^D x \left [{\half} \pa_{\mu} \Phi \pa^{\mu} \Phi  -{m^2 \Phi^2
\over 2}   -\lambda
\Phi^3
\right]
\eqn\twofive
$$
Let us introduce the lightlike coordinates $x^+ ,x^-$ and the transverse
coordinates $X^i$. Light cone quantization is defined by choosing the
direction $x^+$ to play
the role of time. In order to indicate this choice we will relabel the
coordinate $x^+$ and call it $\tau$. The action takes the form
$$
I=\int d\tau dx^- dx^i \left[ 2\pa_{\tau }\Phi\pa_{-}\Phi
-{\half}(\pa_i\Phi\pa_i\Phi
+ m^2 \Phi^2)         -\lambda \Phi^3 \right]
\eqn\twosix
$$
The canonical momentum conjugate to $\Phi$ is given by $2\pa_-  
\Phi$ and the
canonical commutation relations are
$$
[\Phi(X,x^-),\pa_- \Phi(X',{x'}^-)] ={i \over 2} \delta(X-X')\delta(x^- -
{x'}^-)
\eqn\twoseven
$$
Eq \twoseven \ is easily solved by writing $\Phi$ in terms of  
creation and
annihilation operators $a^+(p_-,X),a^-(p_-,X)$. The operator
$a^+(p_-,X)$ creates a particle with longitudinal lightlike  
momentum $p_-$ at
transverse location $X$. The allowable values of $p_-$ are strictly  
positive.
$$
\Phi(X,x_-) \sim \int_0^{\infty}{dp_- \over \sqrt{p_-}}\left[ a^+(p_-,X)
e^{i p_-
x^-}  +  a^-(p_-,X) e^{-i p_-x^-} \right]
\eqn\twoeight
$$
Note that there are no creation or annihilation operators for  
particles of
negative $p_-$. The Fock space is composed of particles of positive
longitudinal momentum from the outset.

The hamiltonian is easily worked out
$$
H=\int_0^{\infty}dp_- \int dX \left({P^2+m^2 \over 2 p_-}\right)
a^+(p_-,X)a^-(p_-,X) + interactions
\eqn\twonine
$$
The interactions conserve longitudinal momentum and are  
transversely local.
Since there are no creation operators for negative $p_-$ it follows  
that the
naive Fock space vacuum is the ground state.

Now it is very easy to define DLCQ for this system. It  consists of
compactifying and periodically identifying the coordinate 
a circle
of radius $R$. The effect is to replace every $p_-$ by a discrete  
variable
$N/R$ with N being integer valued. In order to eventually  pass to the
uncompactified limit we allow $N$ to go to infinity.

 To illustrate the procedure we
will work out the simplest case of 1+1 dimensional real scalar field
theory. First consider
a free field of mass $m$. The Lagrangian is given by
$$
L=\int _0^{ 2 \pi R} dx^- \left[ 2 \pa _{\tau}\Phi \pa _- \Phi  
-{m^2 \over
2} \Phi^2 \right
]
\eqn\twoten
$$
The field is assumed periodic in $x^-$ with period $R$ so that it can be
expanded in a
Fourier series.
$$
\Phi = = \Phi_0 + \sum_{0}^{+ \infty}
\Phi_n e^{i n x^- \over R}+\sum_{0}^{+ \infty}
\Phi^{\dagger}_n e^{-i n x^- \over R}
\eqn\twoeleven
$$
where $\Phi_0$ is the zero momentum mode of $\Phi$.
Inserting eq \twoeleven \ into \ \twoten \ one finds the  
commutation relations
$$
[\Phi_n, \Phi^{\dagger}_m ] = {1\over n} \delta _{nm} \  \ (n \not= 0)
\eqn\twotwelve
$$
In other words the Fock space is composed of quanta with positive  
discrete
longitudinal
momenta. In addition there is a zero mode  $\Phi_0$ whose time derivative
does not enter
the action. When (nonderivative) interactions are added the zero  
mode remains
nondynamical
and may be integrated out. Typical interaction terms such as $V(\Phi)$
induce transitions
between different number of quanta but always in a way that conserves the
integer valued
momentum. The total momentum defines superselection sectors characterized
by an integer
$N$.

In sectors of low $N$ the dynamics is extremely simple. For example the
Fock space vacuum
can not make a transition to any other state. This follows from the
positivity of the
momentum spectrum of the quanta. For the case $N=1$ the dynamics is  
equally
trivial. The
only state with $N=1$ is the state with a single quantum so it too  
can not
mix with any
other state. For $N=2$ there are two states, the state with a single
quantum of 2 units
of momentum and the state with two quanta, each with 1 momentum  
unit. The only
allowed processes
are transitions from 1 to 2 and back. As $N$ increases the number  
of states
and variety
of processes increases and at $N= \infty$ the usual full set of  
light cone
processes can
occur.  However, the DLCQ setup enjoys all of
the advantages of galilean invariance, positivity of longitudinal  
momentum
and simplicity
of vacuum structure not only as $N \to \infty$ but for every finite  
$N$! Of
course for
finite $N$ the breaking of the full Lorentz invariance by the boundary
conditions is felt.

%
%
\chapter{DLCQ of String Theory}

The principles of compactification of string theory are familiar for the
usual situation in which a spacelike coordinate is periodically  
identified.
In this section we will be interested in string theory with periodic
identification of the lightlike direction $x^-$. In other words we  
want to
study the sum over world sheets of arbitrary topology embedded in such a
periodically identified geometry. It is natural to carry this out in a
coordinate system in which $x^-$ is formally treated as a spatial  
coordinate
and $x^+$ is treated as time. The procedure of string quantization  
in such a
light cone frame are well known. The new features required by
compactification of $x^-$ are straightforward.

Compactification by periodic identification generally involves two
modifications. The first is to eliminate all states which are not  
invariant
under translations by $2 \pi R$. In the present case this is simply
accomplished by retaining only the subspace of the string Fock  
space which is
composed of strings carrying lightlike momentum $p_{-} = N/R$. In  
terms of the
usual light cone description of first quantized strings, this means  
that the
usual $\sigma$ coordinate of the world sheet must have length equal  
to $N/R$.
That is, the length of the $\sigma$ axis of any string is  
quantized. Thus a
string of length $2/R $ can split into two strings, each of length  
$1/R$. The
resulting strings can not further split but can rejoin. As in field  
theory the
set of perturbative processes is severely limited in each $N$ sector.
Perturbation theory for
$N=2$ simply consists of repeated splitting and joining transitions  
between
the 1-string and 2-string sectors.

The second and more interesting step is to introduce the twisted sectors
corresponding to strings wound around the periodically identified  
coordinate.
In the present case this means strings satisfying
$$
\int_0^{2\pi R} d{\sigma} {\pa x^{-}(\sigma)\over \pa \sigma} =2\pi \nu R
\eqn\threeone
$$
where $\nu$ is an integer representing the winding number of the string
around $x^-$. What makes this condition different than  
compactification of a
transverse direction $X^i$ is that in light cone quantization, the  
coordinate
$x^{-}$ is not an independent variable. In fact the  constraints of the
theory in light
cone gauge relate ${\pa x^{-}(\sigma)\over \pa \sigma}$ to derivatives of
transverse coordinates.
$$
{\pa x^{-}\over \pa \sigma}={\pa X^{i}\over \pa \sigma}{\pa  
X^{i}\over \pa
\tau} +fermionic \ terms
\eqn\threetwo
$$
Integrating eq \threetwo \ over $\sigma$ and using \threeone \ gives
$$
2\pi \nu R={1 \over p_{-}}(N_L-N_R) ={2 \pi R \over N}(N_L-N_R)
\eqn\threethree
$$
where $N_L,N_R$ are the usual oscillator level numbers of the string.
In other words the usual condition $N_L-N_R = 0$ is replaced by
$$
N_L-N_R =\nu N
\eqn\threefour
$$

In the limit $N \to \infty$ the strings wound around $x^-$ can  
easily be seen
to decouple. For example consider the case $\nu =1$. Eq.  
\threethree can be
written as
$$
(N_L-N_R)=2 \pi R p_-
\eqn\threefive
$$
Furthermore, since $N_L$ and $N_R$ are strictly positive, $N_L+N_R$  
must be
at least of order $p_- R$. Since the light cone energy of a string with
$P_i=0$ is
$$
H={N_L+N_R \over 2 \alpha'  p_-}
\eqn\threesix
$$
we see that the energy is of order $\left({R\over {\alpha'}} \right)$.
On the other hand, the energy of an unwound string is of order ${1\over
\alpha' p_{-}} = {R\over {\alpha' N}}$. Thus in the large $N$ limit  
the energy
of wound strings diverges relative to the energy of strings with  
vanishing
winding number. For finite $N$  and nonvanishing string coupling,   
the wound
strings can be produced in intermediate states even if the total winding
number vanishes.

If one  of the transverse coordinates $X^1$ is compactified eq.
\threefour \ must be modified to include the effects of Kaluza Klein and
winding charges for this direction. If $n$ and $m$ are the integer valued
momentum and winding quantum numbers, eq. \threefour \ is replaced by
$$
N_L-N_R =\nu N +nm
\eqn\threeseven
$$

Interactions can be added to the DLCQ of string theory. As shown by Green
and Seiberg the
form of the interactions is at least in part determined by supersymmetry
and finiteness
of the perturbation series [\gs]. For example the trilinear vertex
for string splitting and joining has the same form as in ordinary  
light cone
string theory except that the strings are restricted to have quantized
longitudinal momentum.
Nothing in the DLCQ setup breaks
either the Galilean symmetry or the spacetime supersymmetry of light cone
string theory. Of course not all symmetries of the large $N$ limit
will be manifested at finite $N$. In particular the symmetries connected
with the Lorentz transformations which rotate the longitudinal  
direction into
transverse directions will be broken by compactifying $x_-$.

One question which will be very important in what follows is the role of
string theory dualities  for the finite $N$ version of the theory. Let us
begin by considering T-duality which interchanges transverse winding and
momentum modes.  . It should be stressed that we are not considering
dualities which act on the longitudinal winding and momentum quantum
numbers.  From eq. \threeseven \ we see that the spectrum of  free  
strings
is invariant under interchange of $n$ and $m$.  Thus the free limit is
manifestly T-dual at finite $N$. Interactions at the tree level do not
change this. For example, in the scattering of ordinary strings T-duality
is satisfied.  To see this note that the tree level amplitudes at  
finite $N$
are identical to the corresponding amplitudes in conventional   
string theory
except for the fact that the external momenta are quantized. A more  
complete
analysis including  loop diagrams will be published elsewhere,  
hopefully by
someone else.  One final point about T-duality is that although it  
has only
been
proved in string perturbation theory it is generally assumed to be  
an exact
nonperturbative symmetry of string theory.

It can also be argued that S-duality should hold in the finite $N$
sectors of the theory. Suppose for example we are studying type IIB  
string
theory. This theory contains both ``fundamental" strings (F-strings) and
D-strings which are related by S-duality. We can imagine building a  
DLCQ based
on extrapolation of weakly interacting F-string perturbation theory  
 or  the
equally valid  extrapolation  of D-string perturbation theory. If  
the results
are not identical there would be two entirely different versions of  
DLCQ for
string theory. Thus the assumption that there exists a unique  
quantization of
string theory in longitudinal compactified spacetime requires  
S-duality to
operate at the level of finite $N$.

%
%
\chapter{DLCQ for M(atrix) Theory}

In this section a stronger form of the conjecture of [\bfss] will be put
forward. The conjecture is this [\motl].

\item{1.}  Discrete light cone quantization of M-theory exists with the
following properties:
\itemitem{a)} The theory exists for every finite value of $N$. Here  
$N$ is to
be interpreted as the total longitudinal momentum $p_-$ in units of  
$1/R$.
\itemitem{b)} Galilean invariance and supersymmetry is manifest for  
all $N$
\itemitem{c)} When the theory is transversely compactified S and  
T-dualities
are
satisfied for all $N$.
\itemitem{d)} In the limit $N \to  \infty$ the theory tends to the usual
M(atrix) description.

\item{2.}  The DLCQ of M-theory is just the finite N version of the  
\sym \
theory used to describe M(atrix) theory in the large $N$ limit. For  
example,
in the case of no compact transverse directions the description is  
the matrix
quantum mechanics of Kabat and Pouliot [\kp]
 and Danielsson, Ferretti and Sundborg [\dfs], describing $N$ D0-branes.
If $K$ directions are compactified on a K-torus the
description is  $K+1$ dimensional \sym with gauge group $U(N)$.

The evidence for the conjecture is both perturbative and nonperturbative.
Let us begin with the perturbative story. Recently a number of  
papers have
demonstrated that weakly coupled type IIA string theory can be derived
[\m] [\bs] [\dvv] from the standard method of compactifying  
M(atrix) theory
[\bfss] [\wati]
[\s].  Let us review the arguments.  The starting point is M(atrix)
theory with one transverse dimension $X^9$ compactified on a circle  
of radius
$L$. (We reserve the symbol $R$ for the longitudinal  
compactification radius).
According to [\wati] the matrix quantum mechanics is replaced by the 1+1
dimensional
\sym \ which is obtained by dimensional reduction of 9+1  
dimensional \sym. The
details of the construction have appeared in several papers and  
will not be
repeated here. We will follow the notations and arguments of ref  
[\dvv] in the
following.
The
\sym
\ has the field content of 8 scalars
$X$, a vector potential and fermionic superpartners. The field theory is
defined
on a periodic spatial coordinate $\sigma$ which varies from $0$ to  
$2\pi$. The
time coordinate is called $\tau$. The action is
$$
S={1\over {2\pi}} \int tr \left((D_{\mu}X^i)^2 +g_s^2 F^2_{\mu \nu}  
-{1\over
g_s^2} [X^i,X^j]^2 + fermionic \ terms \right)
\eqn\fourone
$$
The string coupling $g_s$ is given in terms of the compactification  
radius $L$
by $g_s=L^{3/2}$
 Note that the Yang Mills coupling constant is inversely  
proportional to the
string coupling $g_s$. Thus the Yang Mills theory tends to infinite  
coupling as
$L \to 0$.

In the limit of infinite coupling, the \sym \ is described by a  
super-conformal
fixed point theory [\dvv] [\fhrs]. It is evident from eq \fourone \  
that the
$X's$ must all
commute in this limit. They can therefore be described in terms of  
their $8N$
eigenvalues $X^i_1(\sigma),X^i_2(\sigma),...X^i_N(\sigma)$. The
super-conformal fixed
point theory is identified in [\dvv] as a free field theory of the $X's$
except that the
permutation group acting on the $N$ eigenvalues must be modded out. The
only effect of this
is to modify the periodic boundary conditions on the $X's$. The  
fields $X$ must
be periodic up to a permutation. The result is that the theory has  
disconnected
sectors, each describing a set of noninteracting ``slinkys" [\bs]  
of length
$2\pi N_1,2\pi N_2....2\pi N_n$ where $N_1+N_2+...=N$. Each slinky  
behaves
like a IIA
string and  has exactly the spectrum of the previous section. In  
particular the
individual strings are not restricted to have $N_L=N_R$. The quantization
condition on
$N_L-N_R$ is obtained as follows: The field theoretic (world sheet  
) momentum
is
quantized in integers as a result of the fact that $\sigma$ varies  
from $0$ to
$2\pi$. However, the effective length of the  slinky $(a)$ is $2\pi N_a$.
The effective
quantum of world sheet momentum for that slinky is $1/N_a$. This kind of
fractionation
of quantum numbers is known from black hole physics [\ms]. If the true
field theoretic
momentum is
$\nu$, the slinky carries effective momentum $N_a \nu$ in units of
$1/{N_a}$. Hence
$$
N_L-N_R=N_a \nu
\eqn\fourtwo
$$
This is identical to eq \threeseven. Thus we see that the spectrum of the
strongly coupled  \sym \ with gauge group $U(N)$  coincides with the DLCQ
spectrum of free string theory.

For the interacting theory, it is shown in [\dvv] that the leading
corrections to
free string theory in powers of $g_s$ match the
corrections to the superconformal fixed point theory. The leading  
dimensional
operator causes string splitting and joining in precise  
correspondence with
the type IIA string vertex. Furthermore nothing in this argument uses the
large $N$ limit. Higher order corrections (string genus expansion)  
have not
been calculated either for DLCQ string theory or for \sym  \ but it seems
reasonable that the finiteness and unitarity of each theory will  
force them
to be perturbatively equivalent.

Next we come to the nonperturbative theory. On the string theory  
side there is
no nonperturbative theory but there are many things which are  
believed to be
true. The most important of these are the dualities such as S and T  
duality. T
duality is of course provable in string perturbation theory but the  
real issue
is whether it survives nonperturbative corrections. The recent work  
of the last
two years on duality assumes S and  T-dualities are exact. In fact  
the success
in unifying the different string theories strongly suggests that  the
definition of the nonperturbative theory should include  the  
constraints of
duality.

In the previous section it was explained that the dualities should  
also  be
present in the finite N version of DLCQ of string theory. Thus we  
may assume
that equivalence between finite $N$ M(atrix) theory and DLCQ of  
string theory
requires \sym \  to have certain dualities for all N. Let us review what
is known starting with T-duality. In  [\s], [\grt] the simplest  
constraint
of T-duality was derived for M(atrix) theory compactified on a  
3-torus. It was
shown that T-duality is equivalent to the electromagnetic S-duality  
of 3+1
dimensional \sym. This is one of the oldest known dualities and is  
generally
believed to be exact even for finite $N$. This proves that type IIA  
theory
compactified on a 2-torus is  nonperturbatively self dual under the  
T-duality
which inverts both cycles of the torus and that the T-duality is true for
finite $N$.

A constraint on the behavior of 2+1 dimensional \sym \ follows
from the T-duality which maps type IIA theory to type IIB theory in 9
noncompact dimensions. In [\ss] it was shown that this duality  
requires 2+1
dimensional \sym \ theory to have a strongly coupled fixed point with a
non-manifest $O(8)$ invariance. A proof of this invariance was given in
[\ss] and an independent argument based on the structure of the
superconformal algebra was given by Seiberg [\bs]. Once again the  
arguments
make no
use of the large $N$ limit.

As for S-duality, the simplest example is the self duality of type  
IIB theory.
In the M(atrix) description of IIB theory given in [\ss] the theory is
identified as  2+1  dimensional \sym \ on a 2 torus. In this formulation
S-duality is merely the symmetry under interchange of the 2 cycles of the
torus. It is completely manifest for all $N$.

Evidently then, finite $N$  M(atrix) theory is not only perturbatively
equivalent to string theory but also, in the case of toroidal
compactification,  manifests the nonperturbative dualities that have been
conjectured in recent years. The only properties of the theory  
which require
extrapolation to infinite $N$ are those which involve Lorentz  
transformations
which are explicitly broken by $x^-$ compactification [\p].

Additional nonperturbative evidence for the equivalence of string  
and M(atrix)
theory can be obtained from the spectrum of D-branes. Recall that weakly
coupled string theory has massive D0-branes with mass given by
$$
m^2={1\over g_s^2 \alpha'}
\eqn\fourthree
$$
These excitations  which are nonperturbative can also be found in  
the strongly
coupled 1+1 dimensional \sym. They correspond [\fhrs] [\dvv] to a  
single unit
of
abelian
$U(1)$ electric flux along the compact spatial direction of the  
field theory.

%
%
\chapter{Conclusion}

In this paper, three lines of argument have been related. The first  
is that
string
theory, compactified on a periodically identified lightlike  
coordinate can
be formulated.
This leads to the existence of sectors of the theory characterized by a
quantum number
$N$ defined by $N=p_{-}R$. The limit $N \to \infty$ defines the
uncompactified light
cone quantization of string theory but the finite $N$ version of  
the theory is
a
consistent string theory exhibiting all the dualities of the theory.

The second line of argument is that M(atrix) theory compactified on a
vanishingly small
transverse circle is equivalent to perturbative string theory  
[\motl] [\bs]
[ \dvv].

Finally, the dualities of string theory require related \sym \  
dualities which
surprisingly are satisfied for finite $N$. Together, these point to  
a clear
conclusion.
Finite $N$ M(atrix) theory is the DLCQ of string and M-theory.

In fact all of this points to a way to confirm the conjectured  
equivalence
of the matrix
model and M-theory without ever having to deal with the large $N$  
limit. By
arguments
similar to [\dvv]   we can hope to prove   that all of the weakly coupled
string-theoretic
corners of moduli space are described by an appropriate super Yang Mills
M(atrix) model.
In addition, dualities of
\sym \ would then confirm the nonperturbative connections between the
different corners.
The manifest hamiltonian form of the \sym \ would prove the unitarity of
the theory.
Furthermore all of this can be done without ever having to deal with the
large $N$ limit.

Finally, the connection between \sym \ and DLCQ suggests the  
existence of a
mathematical structure, more comprehensive than ordinary gauge theory. We
do not normally
view each value of total momentum  as defining a separate system.  
The usual
view is
that there is a single dynamical system which can have any value of
momentum. It seems
inevitable that we will eventually think of  the super Yang Mills  
theories
for all $N$ as
being embedded in a single dynamical system. Since in M(atrix) theory we
have generally
identified momentum with gauge flux, it would seem likely that $N$ itself
will be
understood as a flux in some sort of master gauge theory.

\chapter{Acknowledgments}

The  author would like to thank Lubos Motl  and Eric Verlinde for
discussions about the significance of the finite $N$ connection between M
theory and the matrix model.

\refout
\end